\documentclass[a4,12pt,german]{article}

\usepackage{pstricks,pst-node,graphicx}
\usepackage{amssymb,amsmath}

\newtheorem{proposition}{Proposition}[section]
\newtheorem{theorem}{Theorem}[section]
\newtheorem{definition}{Definition}[section]
\newtheorem{corollary}{Corollary}[section]

\newcommand{\be}{\begin{equation}}
\newcommand{\ee}{\end{equation}}
\newcommand{\eea}{\end{eqnarray}}
\newcommand{\bea}{\begin{eqnarray}}

\newcommand{\rx}[1]{{\rm\bf #1}}
\newcommand{\nc}[1]{\langle\langle #1 \rangle\rangle}

\begin{document}


\title{{\bf \Large A noncommutative enumeration problem}}

\author{{\bf Maria Simonetta Bernabei} and
{\bf Horst Thaler} \\[1ex] {Department of Mathematics and Informatics,
University of Camerino,} \\
{Via Madonna delle Carceri 9,
I--62032, Camerino (MC), Italy;}\\
{\small
E-mail:
simona.bernabei@unicam.it, horst.thaler@unicam.it
}}

\date{}

\maketitle

{\abstract In this article we tackle the combinatorics of coloured hard-dimer objects. This is achieved by identifying coloured hard-dimer configurations with a certain class of rooted trees that allow for an algebraic treatment in terms of noncommutative formal power series. \\[2ex]
\\[1ex]
}

\section{Introduction}

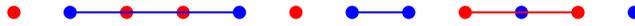
\begin{figure}[h]
\begin{center}
\psset{xunit=1.5cm,yunit=1cm}
\begin{pspicture}(-0.5,-0.5)(7,0)
\psdots[dotstyle=*,linecolor=blue,linewidth=1.5pt](0.5,-0.5)(2.0,-0.5)(3,-0.5)(3.5,-0.5)(4.5,-0.5)(5.5,-0.5)
\psdots[dotstyle=*,linecolor=red,linewidth=1.5pt](0,-0.5)(1,-0.5)(1.5,-0.5)(2.5,-0.5)(4,-0.5)(5,-0.5)
\psline[linecolor=blue](0.5,-0.5)(2,-0.5)
\psline[linecolor=blue](3,-0.5)(3.5,-0.5)
\psline[linecolor=red](4,-0.5)(5,-0.5)
\end{pspicture}
\end{center}
\caption{\label{fig1a} A coloured hard-dimer configuration ($n$=12) with 2 blue dimers, 1 red dimer and 3 inner vertices.}
\end{figure}

The aim is to count coloured hard-dimer configurations which are objects as shown in Figure \ref{fig1a}

More precisely, we define a coloured hard-dimer configuration (CHDC) to be a finite sequence $\sigma_n$, of $n$ blue and red vertices, together with coloured dimers on $\sigma_n$ which must not intersect. Here a coloured dimer is an edge connecting two nearest vertices of the same colour, see Figure \ref{fig1a}. The dimer's colour is given by the colour of its boundary vertices. In graph-theoretic language CHDCs are a subclass of labeled graphs whose vertices and edges carry one of two possible labels. Without loss of generality we may assume that a CHDC is a subset of $\mathbb{R}$ and its vertices belong to $\mathbb{Z}$. We include also the empty CHDC, i.e. the configuration when no dimers are present.

{\it The enumeration problem we have in mind goes as follows: Given a sequence $\sigma_n$ and a triple of nonnegative integers $(i,j,k),$ where $i,j$ count the numbers of blue, red dimers and $k$ the number of inner vertices, i.e. vertices that are connected to their left and right vertex. How many CHDCs of a given type $(i,j,k)$ are there on $\sigma_n$, where $\sigma_n$ as well $n$ may be arbitrary ?}

The motivation for studying this kind of objects is due to their appearance in causally triangulated (2+1)-dimensional gravity. It was shown in \cite{BeLoZa}, by using special triangulations of spacetime, that the discrete Laplace transform of the one-step propagator can be expressed as follows
\begin{equation}\label{eqZ}
Z(u,v,w)=\sum_{n \in \mathbb{N}} e^{-\gamma n} Z_n(u,v,w),
\end{equation}
with
$$
Z_n(u,v,w)=\sum_{\textstyle{\sigma_n}} \displaystyle{\frac{1}{Z_{\sigma_n}^{hcd}(-u,-v,w)}}\;\;\mathrm{and}\;\;
Z_{\sigma_n}^{hcd}(u,v,w)=\sum_{\mathrm{CHDCs}\; D|_{\sigma_n}}u^{|D|_b} v^{|D|_r} w^{|\cap D|}.
$$
In reference \cite{BeLoZa} it is explained how the parameters $\gamma,u,v,w$ are related to physical and geometrical constants.
The exponents $|D|_b,|D|_r,|\hspace{-0.3mm} \cap\hspace{-0.3mm} D|$ count the number of blue dimers, red dimers and inner vertices, respectively. Now, if we want to find explicit expressions for the function $Z$ above, we need to know how many CHDCs, for a given sequence $\sigma_n$, give rise to the same triple of integers $(|D|_b,|D|_r,|\hspace{-0.5mm} \cap\hspace{-0.5mm} D|)$. This is exactly what we are after in this article and we shall find an algebraic solution to the enumeration problem as stated above. The strategy is to identify the set of CHDCs with a certain class of trees the elements of which can be encoded into monomials of a noncommutative formal power series $\sum_{\rx{x}}a_\rx{x}\rx{x}$, where the $\rx{x}$'s are words over the alphabet $\{b,r\}$. This means that the indeterminates $b$ and $r$ are supposed to be noncommutative and a particular word $\rx{x}$ just corresponds to a particular sequence of blue and red vertices $\sigma_n$. The coefficients $a_\rx{x}$ collect the information of CHDCs of a given type on $\sigma_n$.

In section \ref{sec2} we show how to make CHDCs into monomials and thereby solve the enumeration problem from above. In section \ref{sec3} we give an alternative way to express the coefficients $a_{\rx{x}}$ by means of a linear representation of the indeterminates $b,r.$ The formal power series that allow such a representation are called recognizable. This notion is explained in Appendix A and in Appendix B the matrices defining the representation are given. Finally, we use this result to investigate the growth behavior of the number of CHDCs.

\section{From CHDCs to noncommutative series}\label{sec2}
In this section we shall establish a bijection between CHDCs and a certain class of trees. This is achieved by first identifying CHDCs with a particular class of graphs denoted ${\rm\bf G}_{hcd}$ which in turn will be identified with the corresponding class of trees denoted ${\rm\bf T}_{hcd}$.

{\it Bijection {\rm CHDCs} $\Leftrightarrow {\rm\bf G}_{hcd}:$} For a given CHDC we endow its vertices with the order inherited from $\mathbb{Z}$. An extra vertex $\times$ is added in the first place, which becomes the root of the graph. Then all the vertices which are next neighbors are connected through an edge. Finally, the vertices which in the original CHDC are the starting and end points of a dimer are linked via an extra edge to the right, as seen from the root (see Figure \ref{fig1}(a)). We note that only the closest vertices of the same colour may get an extra link and that these become trivalent, unless we consider the last vertex, in which case it becomes bivalent. The set of rooted graphs we obtain in this manner is denoted ${\rm\bf G}_{hcd}$ and it is obvious that we have established a bijection between the latter and the set of CHDCs.

{\it Bijection ${\rm\bf G}_{hcd}\Leftrightarrow {\rm\bf T}_{hcd}:$} When a graph in ${\rm\bf G}_{hcd}$ is given, we move beginning from the root in clockwise direction and cut an edge if the graph remains connected. The cut edge is replaced by two edges ending in univalent vertices called buds (black arrow) and leafs (white arrow), respectively. The procedure is repeated until we reach the last vertex (see Figure \ref{fig1}(b)). We note that buds and three vertices are always connected by one edge only whereas leafs and three vertices may be interlaced by a certain number of bivalent red or blue vertices. The set of rooted trees we obtain following this procedure is denoted ${\rm\bf T}_{hcd}$. Viceversa, when a tree in ${\rm\bf T}_{hcd}$ is given we may, starting from the root, move in clockwise direction and merge the bud leave pairs to edges getting a graph in ${\rm\bf G}_{hcd}$.

We have thus established a bijection between the set of CHDCs and the set ${\rm\bf T}_{hcd}$.
\newcommand{\nr}{\psdots[dotstyle=*,linecolor=red,linewidth=1.5pt](0.0,0.0) }
\newcommand{\tra}{\psset{xunit=0.2cm,yunit=0.2cm}\pspolygon[linewidth=0.5pt](0.35,0.35)(1,-1)(-0.35,-0.35)}
\newcommand{\trb}{\psset{xunit=0.2cm,yunit=0.2cm}\pspolygon*[linewidth=0.5pt](0.35,0.35)(1,-1)(-0.35,-0.35)}
\newcommand{\tri}{\psset{xunit=0.1cm,yunit=0.1cm}\pspolygon[linewidth=0.5pt](0.4,0.9)(-1.8,0.8)(-0.4,-0.9)}
\newcommand{\trf}{\psset{xunit=0.15cm,yunit=0.15cm}\pspolygon*[linewidth=0.5pt,fillstyle=solid](0.5,-0.25)(0.6,1.2)(-0.5,0.25)}
\newcommand{\tria}{\psset{xunit=0.15cm,yunit=0.15cm}\pspolygon[linewidth=0.5pt](0.5,0.25)(-0.6,1.2)(-0.5,-0.25)}
\begin{figure}[t]
\begin{center}
\psset{xunit=1.5cm,yunit=1cm}
\begin{pspicture}(-0.5,0)(7,1)
\psdots[dotstyle=*,linecolor=blue,linewidth=1.5pt](0.5,0.5)(2.0,0.5)(3,0.5)(3.5,0.5)(4.5,0.5)(5.5,0.5)
\psdots[dotstyle=*,linecolor=red,linewidth=1.5pt](0,0.5)(1,0.5)(1.5,0.5)(2.5,0.5)(4,0.5)(5,0.5)
\psline[linecolor=blue](0.5,0.5)(2,0.5)
\psline[linecolor=blue](3,0.5)(3.5,0.5)
\psline[linecolor=red](4,0.5)(5,0.5)
\end{pspicture}
$$\Updownarrow$$
\begin{pspicture}(-0.5,-0.2)(7,1)
\psarc(1.25,0.5){1.13}{180}{360}
\psarc(3.25,0.5){0.38}{180}{360}
\psarc(4.5,0.5){0.75}{180}{360}
\psline(-0.5,0.5)(5.5,0.5)
\psdots[dotstyle=*,linecolor=blue,linewidth=1.5pt](0.5,0.5)(2.0,0.5)(3,0.5)(3.5,0.5)(4.5,0.5)(5.5,0.5)
\psdots[dotstyle=*,linecolor=red,linewidth=1.5pt](0,0.5)(1,0.5)(1.5,0.5)(2.5,0.5)(4,0.5)(5,0.5)
\psdots[dotstyle=x,linewidth=2pt,linewidth=2pt](-0.5,0.5)
\rput(6.5,0.5){(a)}
\end{pspicture}
$$\Updownarrow$$
\begin{pspicture}(-0.5,-0.5)(7,1.4)
\psarc(1.25,0.5){1.13}{180}{360}
\psarc(3.25,0.5){0.38}{180}{360}
\psarc(4.5,0.5){0.75}{180}{360}
\psline(-0.5,0.5)(0.5,0.5)
\psline(0.5,0.5)(0.65,0.8)
\psline(1.1,0.9)(2,0.5)
\psline(2,0.5)(3,0.5)
\psline(3,0.5)(3.1,0.7)
\psline(4,0.5)(4.15,0.8)
\psline(3.5,0.5)(3.4,0.7)
\psline(5,0.5)(4.8,0.9)
\psline(5,0.5)(5.5,0.5)
\psline(3.5,0.5)(4,0.5)
\psdots[dotstyle=*,linecolor=red,linewidth=1.5pt ](1.7,0.63)(1.4,0.76)
\psdots[dotstyle=*,linecolor=blue,linewidth=1.5pt](0.5,0.5)(2.0,0.5)(3,0.5)(3.5,0.5)(4.9,0.72)(5.5,0.5)
\psdots[dotstyle=*,linecolor=red,linewidth=1.5pt](0,0.5)(2.5,0.5)(4,0.5)(5,0.5)
\psdots[dotstyle=x,linewidth=2pt](-0.5,0.5)
\rput(0.65,0.8){\trf}
\rput(1.1,0.9){\tri}
\rput(3.1,0.7){\trf}
\rput(4.15,0.8){\trf}
\rput(3.4,0.7){\tria}
\rput(4.8,0.9){\tria}
\rput(6.5,0.5){(b)}
\end{pspicture}
\caption{\label{fig1} Illustration of the one-to-one correspondence between a coloured hard-dimer, its graph and its tree.}
\end{center}
\end{figure}
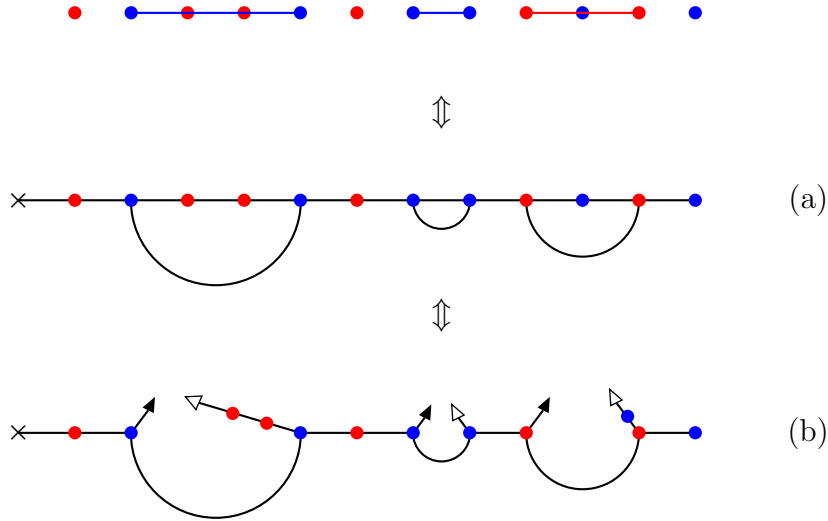
The next goal is to find an algebraic formulation of the enumeration problem for elements in ${\rm\bf T}_{hcd}$. For this we adapt ideas from \cite{BoFrGu} and assign the charges $+1, -1$ and 0 to leaves, buds and univalent coloured vertices, respectively. With this choice any tree has charge 0 and any subtree different from a bud has charge 0 or +1.
For this reason we distinguish between $S$ (respectively, $R$) trees, iff: \\
(a) their total charge is 0 (respectively, 1); \\
(b) any descendant subtree different from a bud has charge 0 or 1. \\
Furthermore, we subdivide these two groups into $S_b$, $R_b$ and $S_r$, $R_r$ trees depending on whether the first vertex after the root is blue or red. \\
Now we are able to write down a system of equations for the four types of trees which will turn out to be linear. Since we are interested in counting the number of CHDCs on a fixed sequence of blue and red vertices we assign noncommutative variables $b$ and $r$ to blue and red vertices, respectively. Furthermore, for a given sequence $\sigma_n$ we group together all CHDCs with the same numbers $|D|_b,|D|_r$ and $|\hspace{-0.5mm}\cap\hspace{-0.5mm} D|$. This can be achieved by encoding a blue or red dimer through the assignment of commutative variables $b_3$ and $r_3$, respectively. Particularly, the variables $b_3$ and $r_3$ are assigned to those trivalent blue and red vertices, respectively, which possess a bud leg pair as subtree. Finally, to every bivalent vertex sitting in between a three vertex and a leaf we assign a commutative variable $y$.

We are thus naturally led to consider the trees $S_b,S_r,R_b,R_r$ as elements of the noncommutative algebra $K\langle\langle b,r \rangle\rangle$ of formal power series where $K$ is the commutative polynomial ring $K=\mathbb{F}[b_3,r_3,y]$ (with $\mathbb{F}$ being any subfield of $\mathbb{R}$ which contains $\mathbb{Z}$). For example, the tree and the corresponding graph in Figure \ref{fig1} read
$$
b_3^2 r_3 y^3\, r\ast b\ast r\ast r\ast b\ast r\ast b\ast b\ast r\ast b\ast r\ast b,
$$
where $\ast$ stands for the noncommutative product in $K\langle\langle b,r \rangle\rangle$.
As we can read off from Figure 3 our trees have to satisfy the following consistency equations:
\begin{eqnarray*}\label{s1}
\hspace{-1cm} S_b &=& b + b\ast (S_b + S_r)+ b_3\, b\ast R_b  \\
\hspace{-1cm} S_r &=& r + r\ast (S_b + S_r)+ r_3\, r\ast R_r  \\
\hspace{-1cm} R_b &=& b+ b\ast (S_b + S_r) + y\, r\ast (1-yr)^{-1}\ast b\ast(S_b + S_r)+ y\, r\ast (1-yr)^{-1}\ast b  \\
\hspace{-1cm} R_r &=& r+ r\ast (S_b + S_r) + y\, b\ast (1-yb)^{-1}\ast r\ast (S_b + S_r) + y\, b\ast (1-yb)^{-1}\ast r 
\end{eqnarray*}

%
%
\noindent
Substituting the expressions of $R_b$ and $R_r$ into the equations for $S_b$ and $S_r$ we get by elementary algebra the following system of equations
\begin{eqnarray*}
& & S_b =\, \underbrace{\left[ b+ b_3\, b^2+ b_3y\, b\ast r\ast (1-y r)^{-1}\ast b\right]}_{=:\, A_b}\ast S_b +  A_b\ast S_r + A_b \\
& & S_r =\, \underbrace{\left[ r+ r_3\, r^2+ r_3y\, r\ast b\ast (1-y b)^{-1}\ast r\right]}_{=:\, A_r}\ast S_r +A_r\ast S_b + A_r,
\end{eqnarray*}
that is
\begin{align}\label{eq1}
 (1-A_b)\ast S_b - A_b \ast S_r &= A_b \nonumber \\
 -A_r\ast S_b +(1-A_r)\ast S_r &= A_r.
\end{align}
The system (\ref{eq1}) can be solved easily with the result
\begin{eqnarray}\label{eqS}
S_b &=&\, \left[1-A_b\ast (1-A_r)^{-1}\right]^{-1}\ast A_b\ast (1-A_r)^{-1} \nonumber \\
&=&\, (1-A_r)\ast (1-A_r-A_b)^{-1} -1.
\end{eqnarray}
By symmetry we also get
$$
S_r =\, (1-A_b)\ast (1-A_r-A_b)^{-1} -1.
$$
\begin{figure}[ht]
\begin{center}
\psset{xunit=1.5cm,yunit=1.5cm}
\begin{pspicture}(-1,-0.5)(1.3,0.5)
\psline(0,0.1)(0,0.5)
\psline[linecolor=green,linewidth=2pt](-0.35,0.5)(0.35,0.5)
\psdots[dotstyle=*,linecolor=blue,linewidth=1.5pt](0.0,0.1)
\rput(0.2,0.1){{\it b}}
\rput(-1,0.0){$S_b=$}
\rput(1,0){+}
\end{pspicture}
\begin{pspicture}(-0.5,-0.5)(1.3,0.5)
\psline(0,0.1)(0,0.5)
\psline(0,0.1)(0,-0.3)
\psline[linecolor=green,linewidth=2pt](-0.35,0.5)(0.35,0.5)
\pscircle[linecolor=blue](0,-0.52){0.34}
\psdots[dotstyle=*,linecolor=blue,linewidth=1.5pt](0.0,0.1)
\rput(0.2,0.1){{\it b}}
\rput(0,-0.97){{\it $S_b$}}
\rput(1,0){+}
\end{pspicture}
\begin{pspicture}(-0.5,-0.5)(1.3,0.5)
\pscircle[linecolor=blue](-0.5,-0.5){0.34}
\psline(-0.35,-0.35)(0,0)
\psline(0,0)(0,0.5)
\psline(0,0)(0.35,-0.35)
\psdots[dotstyle=*,linecolor=blue,linewidth=1.5pt](0.0,0.0)
\psline[linecolor=green,linewidth=2pt](-0.35,0.5)(0.35,0.5)
\rput(0.35,-0.35){\trb}
\rput(-0.2,0.2){{\it $b_3$}}
\rput(0.2,0.2){{\it $b$}}
\rput(-0.5,-0.95){{\it $R_b$}}
\rput(1,0){+}
\end{pspicture}
\begin{pspicture}(-0.5,-0.5)(1.3,0.5)
\psline(0,0.1)(0,0.5)
\psline(0,0.1)(0,-0.3)
\psline[linecolor=green,linewidth=2pt](-0.35,0.5)(0.35,0.5)
\pscircle[linecolor=red](0,-0.52){0.34}
\psdots[dotstyle=*,linecolor=blue,linewidth=1.5pt](0.0,0.1)
\rput(0.2,0.1){{\it b}}
\rput(0,-0.97){{\it $S_r$}}
\end{pspicture}
\end{center}
\begin{center}
\psset{xunit=1.5cm,yunit=1.5cm}
\begin{pspicture}(-1,-0.5)(1.3,1.5)
\psline(0,0.1)(0,0.5)
\psline[linecolor=green,linewidth=2pt](-0.35,0.5)(0.35,0.5)
\psdots[dotstyle=*,linecolor=red,linewidth=1.5pt](0.0,0.1)
\rput(0.2,0.1){{\it r}}
\rput(-1,0){$S_r=$}
\rput(1,0){+}
\end{pspicture}
\begin{pspicture}(-0.5,-0.5)(1.3,0.5)
\psline(0,0.1)(0,0.5)
\psline(0,0.1)(0,-0.3)
\psline[linecolor=green,linewidth=2pt](-0.35,0.5)(0.35,0.5)
\pscircle[linecolor=red](0,-0.52){0.34}
\psdots[dotstyle=*,linecolor=red,linewidth=1.5pt](0.0,0.1)
\rput(0.2,0.1){{\it r}}
\rput(0,-0.97){{\it $S_r$}}
\rput(1,0){+}
\end{pspicture}
\begin{pspicture}(-0.5,-0.5)(1.3,0.5)
\pscircle[linecolor=red](-0.5,-0.5){0.34}
\psline(-0.35,-0.35)(0,0)
\psline(0,0)(0,0.5)
\psline(0,0)(0.35,-0.35)
\psdots[dotstyle=*,linecolor=red,linewidth=1.5pt](0.0,0.0)
\psline[linecolor=green,linewidth=2pt](-0.35,0.5)(0.35,0.5)
\rput(0.35,-0.35){\trb}
\rput(-0.2,0.2){{\it $r_3$}}
\rput(0.2,0.2){{\it $r$}}
\rput(-0.5,-0.95){{\it $R_r$}}
\rput(1,0){+}
\end{pspicture}
\begin{pspicture}(-0.5,-0.5)(1.3,0.5)
\psline(0,0.1)(0,0.5)
\psline(0,0.1)(0,-0.3)
\psline[linecolor=green,linewidth=2pt](-0.35,0.5)(0.35,0.5)
\pscircle[linecolor=blue](0,-0.52){0.34}
\psdots[dotstyle=*,linecolor=red,linewidth=1.5pt](0.0,0.1)
\rput(0.2,0.1){{\it r}}
\rput(0,-0.97){{\it $S_b$}}
\end{pspicture}
\end{center}
\begin{center}
\psset{xunit=1.5cm,yunit=1.5cm}
\begin{pspicture}(-1,-1.5)(2,1.5)
\pscircle[linecolor=blue](-0.5,-0.5){0.34}
\psline(-0.35,-0.35)(0,0)
\psline(0,0)(0,0.5)
\psline(0,0)(0.35,-0.35)
\psdots[dotstyle=*,linecolor=blue,linewidth=1.5pt](0.0,0.0)
\psline[linecolor=green,linewidth=2pt](-0.35,0.5)(0.35,0.5)
\rput(0.35,-0.35){\tra}
\rput(0.2,0.1){{\it $b$}}
\rput(-0.5,-0.95){{\it $S_b$}}
\rput(-0.5,-1.3){(or {\it $S_r$})}
\rput(-1.5,0){$R_b=$}
\rput(1.5,0){+}
\end{pspicture}
\begin{pspicture}(-1,-1.5)(2,1)
\pscircle[linecolor=blue](-0.5,-0.5){0.34}
\psline(-0.35,-0.35)(0,0)
\psline(0,0)(0,0.5)
\psline(0,0)(0.35,-0.35)
\psline[linecolor=green,linewidth=2pt](-0.35,0.5)(0.35,0.5)
\psdots[dotstyle=*,linecolor=blue,linewidth=1.5pt](0.0,0.0)
\psline(0.35,-0.35)(0.7,-0.7)
\psline[linestyle=dashed](0.7,-0.7)(1.05,-1.05)
\psline(1.05,-1.05)(1.4,-1.4)
\multips(0.35,-0.35)(0.35,-0.35){3}{
\nr}
\rput(1.4,-1.4){\tra}
\rput(0.2,0.1){{\it b}}
\rput(0.55,-0.32){{\it r}}
\rput(0.9,-0.64){{\it r}}
\rput(1.26,-1.01){{\it r}}
\rput(0.35,-0.6){{\it y}}
\rput(0.7,-0.95){{\it y}}
\rput(1.05,-1.3){{\it y}}
\rput(-0.5,-0.95){{\it $S_b$}}
\rput(-0.5,-1.3){(or {\it $S_r$})}
\rput(1.5,0){+}
\end{pspicture}
\end{center}
\begin{center}
\psset{xunit=1.5cm,yunit=1.5cm}
\begin{pspicture}(-1,-1.5)(2,0.7)
\psline(0,0)(0,0.5)
\psline(0,0)(0.35,-0.35)
\psline[linecolor=green,linewidth=2pt](-0.35,0.5)(0.35,0.5)
\psdots[dotstyle=*,linecolor=blue,linewidth=1.5pt](0.0,0.0)
\rput(0.35,-0.35){\tra}
\rput(0.2,0.1){{\it b}}
\rput(-1,0){+}
\rput(1.5,0){+}
\end{pspicture}
\begin{pspicture}(-1,-1.5)(2,0.7)
\psline(0,0)(0,0.5)
\psline(0,0)(0.35,-0.35)
\psline[linecolor=green,linewidth=2pt](-0.35,0.5)(0.35,0.5)
\psdots[dotstyle=*,linecolor=blue,linewidth=1.5pt](0.0,0.0)
\psline(0.35,-0.35)(0.7,-0.7)
\psline[linestyle=dashed](0.7,-0.7)(1.05,-1.05)
\psline(1.05,-1.05)(1.4,-1.4)
\multips(0.35,-0.35)(0.35,-0.35){3}{
\nr}
\rput(1.4,-1.4){\tra}
\rput(0.55,-0.32){{\it r}}
\rput(0.9,-0.64){{\it r}}
\rput(1.26,-1.01){{\it r}}
\rput(0.35,-0.6){{\it y}}
\rput(0.7,-0.95){{\it y}}
\rput(1.05,-1.3){{\it y}}
\rput(0.2,0.1){{\it b}}
\end{pspicture}
\caption{\label{fig2} The possible building blocks from the perspective of an arbitrary vertex. The corresponding building blocks for $R_r$ are obtained by exchanging colors and variables ($b \leftrightarrow r $).}
\end{center}
\end{figure}
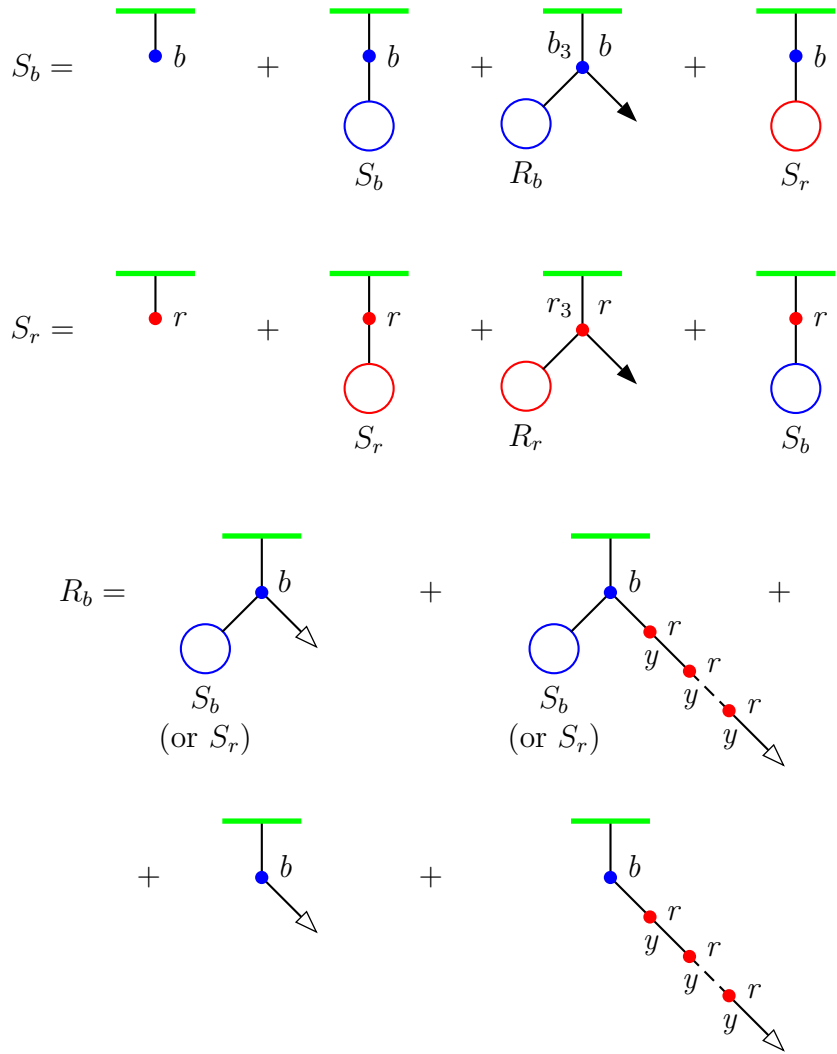
\nopagebreak
\begin{proposition}\label{prop21}
Let $S_b=\sum_{\rx{x}}a_{\rx{x}} \rx{x}$ be the formal series given as part of the unique solution of (\ref{eq1}) in $K\langle\langle b,r\rangle\rangle$. The sum runs over all words $\rx{x}\in \{b,r\}^\ast$ starting with letter $b$ and the coefficients $a_{\rx{x}}=(S_b,\rx{x})$ are given as finite sums
$$
a_{\rx{x}}=\sum_{i,j,k \in \mathbb{N}_0} m_{i,j,k}(\rx{x})\, b_3^i r_3^j y^k,
$$
where the multiplicities $m_{i,j,k}(\rx{x})$ count the number of coloured hard-dimer configurations $D$ on $\rx{x}$ for which $(|D|_b,|D|_r,|\hspace{-0.5mm}\cap\hspace{-0.5mm} D|)=(i,j,k)$. In particular, the evaluation $a_\rx{x}(b_3=1,r_3=1,w=1)$ gives the number of CHDCs on the sequence $\sigma_n$ corresponding to $\rx{x}$. The symmetric assertion holds for $S_r$.
\end{proposition}
{\it Proof.} Let $\#$ be one of the subscripts $b,r$ and decompose $S_\#=\sum_{n=0}^\infty S_\#^n$ where $S_\#^n$ comprises the sum of terms in $S_\#$ with $|\rx{x}|=n$. In the same manner we write $R_\#=\sum_{n=0}^\infty R_\#^n$. An elementary computation shows that $S_\#=\# + \,\{\mbox{higher order terms}\}$ and likewise for $R_\#$, which implies that a solution to (\ref{s1}) gives rise to sequences $(S_\#^n)_{n\geq -1}$ and $(R_\#^n)_{n\geq -1}$ obeying the following recursive equations ($n\geq 1$):
%
\begin{eqnarray}\label{s2}
S_b^n &=& b + b\ast (S_b^{n-1} + S_r^{n-1})+ b_3\, b\ast R_b^{n-1}, \nonumber \\
S_r^n &=& r + r\ast (S_b^{n-1} + S_r^{n-1})+ r_3\, r\ast R_r^{n-1}, \nonumber \\
R_b^n  &=&  b+ b\ast (S_b^{n-1} + S_r^{n-1}) + \sum_{k=1}^{n-2}y^k\, r^k \ast b\ast(S_b^{n-1-k} + S_r^{n-1-k})  \nonumber \\
&& + \, y^{n-1}\, r^{n-1}\ast b, \nonumber \\
R_r^n  &=&  r+ r\ast (S_b^{n-1} + S_r^{n-1}) +\sum_{k=1}^{n-2}y^k\, b^k \ast r\ast(S_b^{n-1-k} + S_r^{n-1-k}) \nonumber \\
&&+\,  y^{n-1}\, b^{n-1}\ast r,
\end{eqnarray}
with initial conditions $S_\#^c = 0, R_\#^c=0$ for $c\in \{-1,0\}$ and $S_\#^1 = R_\#^1=\#.$
Vice versa, for sequences $(S_\#^n)_{n\geq -1}$, $(R_\#^n)_{n\geq -1}$ obeying (\ref{s2}) and related initial conditions, their sums $S_\# :=\sum_{n=-1}^\infty S_\#^n$ and $R_\# :=\sum_{n=-1}^\infty R_\#^n$ will satisfy (\ref{s1}). Therefore we may equivalently look at the system (\ref{s2}). But the elements $S_\#^n$ contain precisely the sum of those terms from $K\langle\langle b,r \rangle\rangle $ which are algebraic counterparts of trees in ${\rm \bf T}_{hcd}$ that have $n$ coloured vertices the first one of which is $\#$. This is due to the fact that there is a one-to-one correspondence between trees from ${\rm \bf T}_{hcd}$ and trees of charge 0 that are constructed recursively according to the building blocks from Figure 3. For a fixed $\rx{x}$, the coefficient $a_{\rx{x}}$ is a sum of elements in $K$ which stem from trees that are constructable for the particular sequence of blue and red vertices given by $\rx{x}.$ Since the indeterminates $b_3, r_3, y $ are commutative, trees will contribute the same term $b_3^i r_3^j y^k$, whenever the numbers of vertices corresponding to a $b_3, r_3, y$ indeterminate, coincide. Moreover, a glance at the bijection $\rm{CHDCs} \Leftrightarrow \mathbf{T}_{{\it hcd}}$ shows that $(i,j,k)= (|D|_b,|D|_r,|\hspace{-0.5mm}\cap\hspace{-0.5mm} D|)$, which proves the assertion. $\qquad \Box$ \\[1ex]
As immediate consequence we get
\begin{corollary}
Let $S_b=\sum_{\rx{x}}a_{\rx{x}} \rx{x}$ be the formal power series representing the solution of (\ref{eq1}). The discrete Laplace transform from (\ref{eqZ}) can be expressed, using Proposition \ref{prop21}, by means of
$$Z^{hcd}_{\sigma_n}(u,v,w)=[(a_{\rx{x}}(b_3=u,r_3=v,y=w)+a_\rx{x}(b_3=v,r_3=u,y=w)].$$
In the latter equality we have evaluated the coefficients $a_{\rx{x}}$ at points $u,v,w\in \mathbb{C}$ and have exploited the symmetric relation between $S_b$ and $S_r$.
\end{corollary}

\section{The solution $S_b$ as a recognizable series}\label{sec3}
In order to find the coefficients in $S_b$ we might start directly from (\ref{eqS}). This, however, necessitates to invert and multiply noncommutative power series. A recursive procedure or the use of Cauchy's product would mean an exponentially growing number of operations. Therefore, if we'd like to employ numerical or symbolical programs on computers,
it is worthwhile making some more effort to find an explicit representation for the coefficients $a_{\rx{x}}$ of $S_b$ by means of matrices. This is in fact possible as $S_b$ turns out to be a recognizable series, see Appendix A.
\begin{proposition}\label{prop31}
The series $S_b$ is a recognizable series with representation
$$\mu_1:\{b,r\}^\ast \rightarrow K^{19\times 19}$$ given by $\mu_1(b)=B_1,\mu_1(r)=R_1$, where the matrices $B_1,R_1$ and tuples $\lambda_1,\gamma_1\in K^{19},$ are given in Appendix B. An analogous statement holds for $S_r$ with $\mu_2(r)=B_1(b_3 \mapsto r_3)$ and $\mu_2(b)=R_1(r_3 \mapsto b_3).$ Since the property of being recognizable is preserved under algebraic operations we also have that $S_b+S_r$ is recognizable.
\end{proposition}
{\it Proof.} By Proposition \ref{propa1} we have to show that $S_b$ is contained in a finitely generated (f.g.) stable $K$-submodule of $K\langle\langle b,r\rangle\rangle.$ Hence we shall verify one by one that the following elements are recognizable series: $A_r, 1-A_r-A_b$ and $S_b$. Two basic rules are needed for proving the statement. For a letter $a$ and any two formal series $P,Q$ we have:
$$a^{-1}(PQ)=(a^{-1}P)Q+(P,1)(a^{-1}Q)$$
and for a proper series $Q$
$$a^{-1}(Q^\ast)=(a^{-1}Q)Q^\ast.$$
So first let's have a look at
$$A_r=r+ r_3\, r^2+ r_3y\, r\ast b\ast (1-y b)^{-1}\ast r.$$
It is not difficult to see that the finitely generated (f.g.) $K$-submodule
$$
U_1:=K 1 \oplus K S_2 \oplus K r \oplus K S_2\ast r \oplus K b \oplus Kb\ast S_2 \oplus Kb\ast S_2\ast r,$$
where $S_2=(1-y b)^{-1}$, is stable and contains the term $ b\ast (1-y b)^{-1}\ast r$. In the same vein one shows that the $K$-submodule $U_2$ generated by
$$S_1= 1,\, S_2=(1-y b)^{-1},\, S_3= r,\, S_4=S_2\ast r,\, S_5= b,\, S_6= b\ast S_2,\, S_7= b\ast S_2\ast r,$$
$$
S_8= r\ast b,\, S_9= r\ast b\ast S_2,\, S_{10}= r\ast b\ast S_2\ast r,\, S_{11}= r^2.$$
is stable, f.g. and contains $A_r = S_3 + r_3 S_{11}+r_3 y\, S_{10}.$ \\
Enlarging $U_2$ by the following basis elements
$$S_{12}= (1-y r)^{-1},\, S_{13}= S_{12}\ast b,\, S_{14}= r\ast S_{12},\,S_{15} = r\ast S_{12}\ast b,\, S_{16} = b\ast r,
$$
$$
S_{17} =b\ast r\ast S_{12},\, S_{18} =b\ast r\ast S_{12}\ast b,\, S_{19}=b^2
$$
we get a f.g. stable $K$-submodule $U_3$ which contains $1-A_r$ and
$$1-A_r-A_b=1-S_3-r_3\, S_{11}-r_3y\, S_{10}-S_5-b_3\, S_{19}-b_3 y\, S_{18}.$$
Finally, we define $P:= A_r+A_b$, $P^\ast := (A_r+A_b)^\ast$ and the $K$-submodule $U_4$ generated by the basis elements
$$T_1:=1,\, T_2 =P^\ast,\, T_3:=S_2\ast P^\ast,\, T_4:=S_3\ast P^\ast,\ldots ,\, T_{20}:=S_{19}\ast P^\ast.
$$
An elementary, yet tedious calculation shows again that $U_4$ is a f.g. stable submodule containing
$$
S_b=(1-A_r)\ast (1-A_r-A_b)^{-1}-1=(1-A_r)\ast P^\ast -1
$$
$$
=P^\ast -S_3\ast P^\ast - r_3 S_{11}\ast P^\ast -r_3y\, S_{10}\ast P^\ast -1.
$$
The corresponding statement for $S_r$ follows once more from symmetry considerations.\hfill \\[1ex]
$\Box $

The result above allows to study the asymptotic behavior of the number of CHDCs on $\sigma_n$ when $n$ tends to infinity. According to Proposition (\ref{prop31}) $S_b+S_r$ is a recognizable series and the evaluation at $b_3=r_3=y=1$ again gives a recognizable series. This entails that for
$$M= S_b(b_3=1,r_3=1,y=1)+S_r(b_3=1,r_3=1,y=1)$$ there exists a representation $\mu':\{b,r\}^\ast \rightarrow \mathbb{R}^{d\times d}$ and vectors $\lambda',\gamma'\in \mathbb{R}^{d}$ such that
$(M,\rx{x})=\lambda'^\top \cdot \mu'(\rx{x})\cdot \gamma'$. This representation is given by
$$\mu'(b)=B(b_3=1,y=1)\;\;\text{and}\;\; \mu'(r)=R(r_3=1,y=1)$$ and the vectors by $$\lambda'=\lambda(b_3=1,r_3=1,y=1),\;\gamma'=\gamma,$$ see Appendix B. Note that $(M,\rx{x})$ counts the number of CHDCs on the sequence of blue and red sites corresponding to $\rx{x}$, irrespective of whether this sequence starts with $b$ or $r$.

It is convenient to discuss the asymptotics of CHDCs within the framework of ergodic dynamical systems. To be precise, let $B':=\mu'(b),R':=\mu'(r)$ and define $\nu$ as the probability measure on $\{B',R'\}$ given by $\nu(B')=\nu(R')=\frac{1}{2}.$ Then as ergodic dynamical system $(\Omega,\mathcal{F},\mathbb{P},\theta)$ we choose

\begin{itemize}
\item $\Omega=\{B',R'\}^\mathbb{N}$, i.e. the set of sequences $\omega=(\omega_1,\omega_2,\ldots)$ with $\omega_i \in \{B',R'\}.$ \newline
    If $g_n(\omega)=\omega_n$ denotes the coordinate maps then $\mathcal{F}$ should be the $\sigma$-algebra generated by the $g_n$'s.
\item $\mathbb{P}$ is the product measure $\mathbb{P}=\prod_{n=1}^\infty \nu.$
\item $\theta$ is the shift operator $\theta(\omega)=(\omega_2,\omega_3,\ldots).$
\end{itemize}
Upon turning $\{b,r\}^\mathbb{N}$ into a probability space that is isomorphic to $(\Omega,\mathcal{F},\mathbb{P})$ we may state the following theorem
\begin{theorem}
For $\omega\in \{b,r\}^\mathbb{N}$ let $\omega_n:=(\omega_1,\ldots,\omega_n)$ and let
$D_n(\omega)$ denote the number of CHDCs on $\omega_n$.
There exists a finite constant random variable $\alpha$ on $(\{b,r\}^\mathbb{N},\mathcal{F},\mathbb{P)}$ such that for almost all $\omega\in \{b,r\}^\mathbb{N}$
$$\frac{1}{n}\ln D_n(\omega)\rightarrow \alpha,\quad \mbox{as}\;\; n\rightarrow \infty.$$
This means that the number of CHDCs grows exponentially and that asymptotically this growth rate is the same for almost all $\omega$.
\end{theorem}
{\it Proof.}
If we define the matrix-valued random variables by
$$X_n(\omega)=g_1(\omega)g_2(\omega)\cdots g_{n}(\omega),$$
then
\begin{equation} \label{eqYn}
Y_n=-\ln \left(\lambda'^\top\cdot X_n \cdot \gamma'\right),
\end{equation}
defines a subadditive sequence of random variables, i.e.
$$Y_{m+n}\leq Y_m\circ \theta^n +Y_n, \quad\text{for all $m$ and $n$}.$$ To see subadditivity we first note that with the natural identifications $\omega_n \leftrightarrow \rx{x}:(\omega_1,\ldots,\omega_n)= \omega_1\ast \cdots \ast \omega_n=\rx{x}$ we may write $\lambda'^\top\cdot X_n(\omega) \cdot \gamma'=(M,\rx{x}).$ Writing $\rx{x}=\rx{x}_m\ast\rx{x}_{m'}$, with $|\rx{x}_m|=m$ and $|\rx{x}_{m'}|=m'$, subadditivity follows from the combinatorial fact that $(M,\rx{x})\geq (M,\rx{x}_m)(M,\rx{x}_{m'}).$
Now Kingman's subadditive ergodic theorem, see \cite{Kin} and \cite[Th. IV.1.2]{CaLa}, applied to our ergodic dynamical system $(\Omega,\mathcal{F},\mathbb{P},\theta)$ implies that there exists a random variable $\beta$ such that
$$\lim_{n\rightarrow \infty}\left(\frac{Y_n}{n}\right)= \beta=\mathbb{E}[\beta],\quad \mathbb{P}\mbox{-a.s.}.$$
In addition one has $\beta=\inf_{n}\left(\mathbb{E}[n^{-1}Y_n]\right)$.
A priori it is not clear whether $\beta > -\infty
$ which we clarify now. By Jensen's inequality we find
\begin{equation}\label{eqJen}
\mathbb{E}\left[n^{-1}Y_n\right]\geq -n^{-1}\ln(\lambda'^\top\cdot \mathbb{E}[X_n]\cdot \gamma')= -n^{-1}\ln(\lambda'^\top\cdot Z^n\cdot \gamma'),
\end{equation}
where $Z=\dfrac{B'+R'}{2}$. From the very definitions we deduce easily that
$$\lambda'^\top\cdot Z^n\cdot \gamma'=2\lambda'^\top\cdot \left(\frac{B_1(b_3=1,y=1)+R_1(r_3=1,y=1)}{2}\right)
^n\cdot \gamma'=:f(n).$$
An explicit calculation shows that the dominant eigenvalue of the nonnegative matrix
$$\Xi:=\frac{B_1(b_3=1,y=1)+R_1(r_3=1,y=1)}{2}$$
is $3/2$ with algebraic multiplicity one. Using an appropriate representation for the matrix powers $\Xi^n$ like that employed in \cite[Th. 8, Ch. III]{Gan} shows that
$\lim_{n\rightarrow \infty}-n^{-1}\ln f(n)=-\dfrac{3}{2}.$ This in turn implies $\mathbb{E}\left[n^{-1}Y_n\right]\geq \text{const.},\;\forall\, n,$ for some
$n$-independent finite constant and consequently $\beta=\inf_n(\mathbb{E}[n^{-1}Y_n])\geq \text{const.}$ We have thus established almost sure convergence of
$$\frac{1}{n}\ln D_n=-\frac{Y_n}{n}\rightarrow -\beta =: \alpha,\quad \mbox{as}\;\; n\rightarrow \infty. \qquad \Box$$
\begin{appendix}
\section{Formal power series}
In this Appendix we recall some basics related to recognizable power series. For more details the reader is referred to \cite{BeReu} or \cite{Sa}.
Suppose we are given a semiring $K$ and an alphabet $A$ with its associated free monoid $A^\ast$ and unit $1$.
A {\it formal noncommutative power series} $S$ or simply a {\it formal series} is a map
$$S :A^\ast \rightarrow K.$$
The image by $S$ of a word $\rx{x}$ is denoted by $(S,\rx{x})$ and is called {\it coefficient} of $\rx{x}$ in $S$. The coefficient $(S,1)$ is called {\it constant term} and a formal series $S$ whose constant term is zero is called {\it proper}. For $\rx{x}=x_1 \ast\cdots\ast x_n$ we define $|\rx{x}|=n$ to denote the length of the word $\rx{x}$ with the convention $|1|=0$. The set of formal series with coefficients in $K$ is denoted by $K \nc{A} $.
We shall consider $K \nc{A} $ as being equipped with the natural semiring structure where addition and product are given by
$$ (S+T)(\rx{x}):=(S,\rx{x})+(T,\rx{x}),\quad (ST)(\rx{x}):=\sum_{\rx{y},\,\rx{z}\in A^\ast :\,\rx{y}\ast\rx{z}=\rx{x}}(S,\rx{y})(T,\rx{z}). $$
Moreover, there are left and right external operations of $K$ on $K \nc{A} $ which are given by
$$(k S,\rx{x}):=k(S,\rx{x}); \quad (Sk,\rx{x}):=(S,\rx{x})k.$$
The subalgebra $K\langle A\rangle$ is defined to comprise those formal series with only finitely many terms different from zero.
It is obvious that with the operations defined above $K \nc{A} $ becomes an algebra when $K$ is a ring. The semiring $K \nc{A}$ can easily be turned into a complete topological semiring. This is achieved by introducing the following metric. First, let
$$\kappa:
K \nc{A} \times K \nc{A}  \rightarrow \mathbb{N}_0 \cup \infty,$$
$$
\kappa(S,T):=\inf\{n\in \mathbb{N}_0\, |\, \exists \, \rx{x}\in A^\ast : |\rx{x} |=n\; \mbox{and}\; (S,\rx{x})\neq (T,\rx{x})\}
$$
and then define the metric
$$ d: K \nc{A}\times K \nc{A}\rightarrow \mathbb{R}_+, $$
$$
d(S,T):= \frac{1}{2^{\kappa(S,T)}}.
$$
It is easy to see that with respect to this topology a sequence $(S_n)_{n\in\mathbb{N}}$ converges to $S$ if and only if $\forall \, m \in\mathbb{N}_0 \,\exists\, n_0\in \mathbb{N}:\left[\forall \, n \geq n_0 \;\mbox{and}\; \forall\, \rx{x} \;\mbox{with}\; |\rx{x}|\leq m :(S_n,\rx{x})=(S,\rx{x})\right].$ This fact then implies that every Cauchy sequence is convergent and it becomes natural to write elements $S\in  K \nc{A}$ in the form
$$
S =\sum_{\rx{x}\in A^\ast }(S,\rx{x})\rx{x}.
$$
For a proper series $S$ it is possible to define the {\it star operation} by $S^\ast := \sum_{n \geq 0}^\infty S^n.$
\begin{definition} \label{def1App}
{\rm (a)}
The \textsf{rational operations} in $K\nc{A}$ are the sum, product, the two external operations and the star operation. A subset of $K\nc{A}$ is rationally closed if it is closed with respect to the rational operations. The \textsf{rational closure} of a subset $E$ of $K\nc{A}$ is the intersection of all rationally closed subsets of $K\nc{A}$ which contain $E.$ When $K$ is a ring the star operation and inversion play the same role since $S^\ast =(1-S)^{-1}.$ \\
{\rm (b)} A formal series is called \textsf{rational} if it is contained in the rational closure of $K\langle A\rangle.$ \\
{\rm (c)} A formal series $S$ is called \textsf{recognizable} if there is some $n\geq 1$, a homomorphism of semirings (or simply a representation)
$$
\mu :A^\ast \rightarrow K^{n\times n},
$$
where $K^{n\times n}$ carries its multiplicative structure and tuples $\lambda, \gamma\in K^{n},$ such that for all words $\rx{x}$
$$
(S,\rx{x})=\lambda^\top\cdot \mu(\rx{x})\cdot \gamma .
$$
\end{definition}
A $K$-linear operation of $A^\ast$ on $K\nc{A}$ can be defined by setting
$$
\rx{y}^{-1}S:=\sum_{\rx{x}\in A^\ast}(S,\rx{y}\ast\rx{x})\rx{x}, \quad \rx{y}\in A^\ast.
$$
A subset $M$ of $K\langle\langle A \rangle\rangle $ is called {\it stable} if for all $\rx{x}\in A^\ast$ we have $\rx{x}^{-1}(M)\subset M.$
The following proposition is the core for proving Proposition \ref{prop31}
\begin{proposition}\label{propa1}
A formal series $S$ is recognizable if and only if there exists a stable finitely generated left $K$-submodule of $K\nc{A}$ which contains $S.$
\end{proposition}
In addition the following equivalence holds
\begin{theorem}\label{thschue} {\rm (}Sch\"utzenberger{\rm)} A formal series is recognizable if and only if it is rational.
\end{theorem}

\section{The matrices $B$ and $R$}
Let $(T_1,\ldots,T_{20})$ be the basis elements spanning the submodule $U_4$ as defined in the proof of Proposition \ref{prop31}. Furthermore, let $\widetilde{B}=:\mu_1(b),\widetilde{R}=:\mu_1(r)$ be the matrices given by $b^{-1}T_i=\sum_{j=1}^{20} \widetilde{B}_{ij}T_j$ and $r^{-1}T_i=\sum_{j=1}^{20} \widetilde{R}_{ij}T_j.$ It turns out that the first row and first column of $\widetilde{B}$ and $\widetilde{R}$ contain only zeros which implies that we may reduce the representation $\mu_1$ to the subspace $V_1:=\text{span}_K\{T_2,\ldots,T_{20}\}$ where it is determined by $\mu_1(b)=B_1,\mu_1(r)=R_1$ as \\[2ex]
$$
B_1=\left(
  \begin{array}{ccccccccccccccccccc}
      1 & 0 & 0 & 0 & b_3 & 0 & 0 & 0 & 0 & 0 & 0 & 0 & 0 & 0 & b_3 y & 0 & 0 & 0 & 0 \\
     1 & y & 0 & 0 & b_3 & 0 & 0 & 0 & 0 & 0 & 0 & 0 & 0 & 0 & b_3 y & 0 & 0 & 0 & 0 \\
     0 & 0 & 0 & 0 & 0 & 0 & 0 & 0 & 0 & 0 & 0 & 0 & 0 & 0 & 0 & 0 & 0 & 0 & 0 \\
     0 & 0 & 0 & y & 0 & 0 & 0 & 0 & 0 & 0 & 0 & 0 & 0 & 0 & 0 & 0 & 0 & 0 & 0 \\
     1 & 0 & 0 & 0 & 0 & 0 & 0 & 0 & 0 & 0 & 0 & 0 & 0 & 0 & 0 & 0 & 0 & 0 & 0 \\
     0 & 1 & 0 & 0 & 0 & 0 & 0 & 0 & 0 & 0 & 0 & 0 & 0 & 0 & 0 & 0 & 0 & 0 & 0 \\
     0 & 0 & 0 & 1 & 0 & 0 & 0 & 0 & 0 & 0 & 0 & 0 & 0 & 0 & 0 & 0 & 0 & 0 & 0 \\
      \multicolumn{19}{c}{\mathbf{0}^{4\times 19}} \\
     1 & 0 & 0 & 0 & b_3 & 0 & 0 & 0 & 0 & 0 & 0 & 0 & 0 & 0 & b_3 y & 0 & 0 & 0 & 0 \\
     1 & 0 & 0 & 0 & 0 & 0 & 0 & 0 & 0 & 0 & 0 & 0 & 0 & 0 & 0 & 0 & 0 & 0 & 0 \\
     \multicolumn{19}{c}{\mathbf{0}^{2\times 19}} \\
     0 & 0 & 1 & 0 & 0 & 0 & 0 & 0 & 0 & 0 & 0 & 0 & 0 & 0 & 0 & 0 & 0 & 0 & 0 \\
     0 & 0 & 0 & 0 & 0 & 0 & 0 & 0 & 0 & 0 & 0 & 0 & 0 & 1 & 0 & 0 & 0 & 0 & 0 \\
     0 & 0 & 0 & 0 & 0 & 0 & 0 & 0 & 0 & 0 & 0 & 0 & 0 & 0 & 1 & 0 & 0 & 0 & 0 \\
     0 & 0 & 0 & 0 & 1 & 0 & 0 & 0 & 0 & 0 & 0 & 0 & 0 & 0 & 0 & 0 & 0 & 0 & 0 \\

  \end{array}
\right)
$$
\\[2ex]
$$R_1=\left(
  \begin{array}{ccccccccccccccccccc}

     1 & 0 & r_3 & 0 & 0 & 0 & r_3 y & 0 & 0 & 0 & 0 & 0 & 0 & 0 & 0 & 0 & 0 & 0 & 0 \\
     1 & 0 & r_3 & 0 & 0 & 0 & r_3 y & 0 & 0 & 0 & 0 & 0 & 0 & 0 & 0 & 0 & 0 & 0 & 0 \\
     1 & 0 & 0 & 0 & 0 & 0 & 0 & 0 & 0 & 0 & 0 & 0 & 0 & 0 & 0 & 0 & 0 & 0 & 0 \\
     1 & 0 & 0 & 0 & 0 & 0 & 0 & 0 & 0 & 0 & 0 & 0 & 0 & 0 & 0 & 0 & 0 & 0 & 0 \\
     \multicolumn{19}{c}{\mathbf{0}^{3\times 19}} \\
     0 & 0 & 0 & 0 & 1 & 0 & 0 & 0 & 0 & 0 & 0 & 0 & 0 & 0 & 0 & 0 & 0 & 0 & 0 \\
     0 & 0 & 0 & 0 & 0 & 1 & 0 & 0 & 0 & 0 & 0 & 0 & 0 & 0 & 0 & 0 & 0 & 0 & 0 \\
     0 & 0 & 0 & 0 & 0 & 0 & 1 & 0 & 0 & 0 & 0 & 0 & 0 & 0 & 0 & 0 & 0 & 0 & 0 \\
     0 & 0 & 1 & 0 & 0 & 0 & 0 & 0 & 0 & 0 & 0 & 0 & 0 & 0 & 0 & 0 & 0 & 0 & 0 \\
     1 & 0 & r_3 & 0 & 0 & 0 & r_3 y & 0 & 0 & 0 & 0 & y & 0 & 0 & 0 & 0 & 0 & 0 & 0 \\
     0 & 0 & 0 & 0 & 0 & 0 & 0 & 0 & 0 & 0 & 0 & 0 & y & 0 & 0 & 0 & 0 & 0 & 0 \\
     0 & 0 & 0 & 0 & 0 & 0 & 0 & 0 & 0 & 0 & 0 & 1 & 0 & 0 & 0 & 0 & 0 & 0 & 0 \\
     0 & 0 & 0 & 0 & 0 & 0 & 0 & 0 & 0 & 0 & 0 & 0 & 1 & 0 & 0 & 0 & 0 & 0 & 0 \\
     \multicolumn{19}{c}{\mathbf{0}^{4\times 19}} \\

\end{array}
\right)
$$
\\[1ex]
Finally, define $\widetilde{\lambda} ,\widetilde{\gamma}$ by $S_b=\sum_{i=1}^{20}\widetilde{\lambda}_i T_i$ and $\widetilde{\gamma}_i=(T_i,1)$ and set $$\lambda_1 :=(\widetilde{\lambda}_2,\widetilde{\lambda}_3,\ldots,\widetilde{\lambda}_{20}),\;\gamma_1 :=(\widetilde{\gamma}_2,\widetilde{\gamma}_3,\ldots,\widetilde{\gamma}_{20}),$$ then \\
$\lambda_1 = \left(
             \begin{array}{ccccccccccccccccccc}
               1 & 0 & -1 & 0 & 0 & 0 & 0 & 0 & 0 & -r_3 y & -r_3 & 0 & 0 & 0 & 0 & 0 & 0 & 0 & 0
             \end{array}
           \right),
$
\\[1ex]
$\gamma_1 = \left(
             \begin{array}{ccccccccccccccccccc}
               1 & 1 & 0 & 0 & 0 & 0 & 0 & 0 & 0 & 0 & 0 & 1 & 0 & 0 & 0 & 0 & 0 & 0 & 0
             \end{array}
           \right).\\[1ex]
$

The corresponding representation for $S_r$ can be found easily by exchanging the letters $b$ and $r$. For this we define the new basis elements
$$\widehat{T}_i:=T_i(b \leftrightarrow r)_{2\leq i\leq 20},$$ which are linearly independent of the $T_i$'s, and set $V_2:=\text{span}_K\{\widehat{T}_2,\ldots,\widehat{T}_{20}\}$. \newline Furthermore we put
$\lambda_2:=\lambda_1(r_3\mapsto b_3)$ and $\gamma_2:=\gamma_1.$ Thus the series $S_r$ is recognizable with representation
$$\mu_2:\{b,r\}^\ast\rightarrow K^{19\times 19},$$ given by $$\mu_2(b)=B_2:=R_1(r_3\rightarrow b_3),\;\mu_2(r)=R_2:=B_1(b_3\rightarrow r_3),$$ and tuples $\lambda_2,\gamma_2\in K^{19}.$ It is readily seen that also $S_b+S_r$ is recognizable with its representation defined on $V_1\oplus V_2$:
$$\mu:\{b,r\}^\ast \rightarrow K^{38\times 38}$$ is specified by $$\mu(b)=B:=
\left(
  \begin{array}{cc}
    B_1 & \textbf{0} \\
    \textbf{0} & B_2 \\
  \end{array}
\right)\;\; \text{and}\;\;
\mu(r)=R:=
\left(
  \begin{array}{cc}
    R_1 & \textbf{0} \\
    \textbf{0} & R_2 \\
  \end{array}
\right),
$$
and the tuples $\lambda,\gamma\in K^{19}\oplus K^{19}$ are given as
$\lambda=(\lambda_1,\lambda_2)$ and $\gamma=(\gamma_1,\gamma_2).$
\end{appendix}


\begin{thebibliography}{999999}

\bibitem{BeLoZa} Benedetti, D., Loll, R., Zamponi, F.: $(2+1)$-dimensional quantum gravity as the continuum limit of causal dynamical triangulations.  Phys. Rev. D {\bf 76}, no. 10, 104022, 26 pp. (2007).

\bibitem{BeReu} Berstel, J., Reutenauer, C.: Rational Series and their Languages. EATCS Monographs on Theoretical Computer Science, 12. Springer-Verlag, Berlin (1988).

\bibitem{BoFrGu} Bouttier, J., Di Francesco, P., Guitter, E.: Census of planar maps: from the one-matrix model solution to a combinatorial proof.
Nuclear Phys. B {\bf 645}, no. 3, 477-499 (2002).

\bibitem{CaLa} Carmona, R., Lacroix, J.: Spectral Theory of Random Schr\"odinger Operators. Birkh\"auser, Boston (1990).

\bibitem{Gan} Gantmacher, F. R.: Applications of the Theory of Matrices. Interscience-Publishers, Inc., New York (1959).

\bibitem{Kin} Kingman, J. F. C.: The ergodic theory of subadditive stochastic processes.  J. Roy. Statist. Soc. Ser. B {\bf 30}, 499--510 (1968).

\bibitem{Sa} Salomaa, A., Soittola, M.: Automata-theoretic Aspects of Formal Power Series. Texts and Monographs in Computer Science. Springer-Verlag, New York-Heidelberg (1978).

\bibitem{Sen} Seneta, E.: Non-negative Matrices and Markov chains. Springer-Verlag, New York (2006).

\end{thebibliography}
\end{document}